\documentclass[preprint,review,12pt]{elsarticle}

\makeatletter
\def\ps@pprintTitle{%
 \let\@oddhead\@empty
 \let\@evenhead\@empty
 \def\@oddfoot{\centerline{\thepage}}%
 \let\@evenfoot\@oddfoot}
\makeatother

\usepackage{graphicx}  
\usepackage{dcolumn}  
\usepackage{bm}           
\usepackage{amsmath}
\usepackage{mathrsfs}
\usepackage{float}
\usepackage{listings}
\usepackage{multirow}
\usepackage{subfigure}
\usepackage{amsfonts}

\begin{document}

\begin{frontmatter}

%
%

\title{Continuous Time Random Walks
for Non-Local Radial Solute Transport
}

%
%

\author{Marco Dentz}
\address{Institute of Environmental Assessment and Water Research
  (ID\mbox{\AE}A), Spanish National Research Council (CSIC),
08034 Barcelona, Spain}

\author{Peter K. Kang}
\address{Massachusetts Institute of Technology, 77
  Massachusetts Ave, Building 48, Cambridge, Massachusetts 02139, USA}

\author{Tanguy Le Borgne}
\address{Universit\'e de Rennes 1, CNRS, Geosciences Rennes,
  UMR 6118, Rennes, France}

\date{\today}
\begin{abstract}
This paper derives and analyzes continuous time random walk (CTRW) models 
in radial flow geometries for the quantification of non-local solute transport  
induced by heterogeneous flow distributions and by mobile-immobile mass transfer processes.
To this end we derive a general CTRW framework in radial
coordinates starting from the random walk equations for radial particle
positions and times. The particle density, or solute
concentration is governed by a non-local radial advection-dispersion
equation (ADE). Unlike in CTRWs for uniform flow scenarios, 
particle transition times here depend on the radial
particle position, which renders the CTRW non-stationary. 
As a consequence, the memory kernel characterizing the non-local ADE,
is radially dependent. Based on this general formulation, we derive
radial CTRW implementations that (i) emulate non-local radial transport due to
heterogeneous advection, (ii) model multirate mass transfer (MRMT)
between mobile and immobile continua, and (iii) quantify both
heterogeneous advection in a mobile region and mass
transfer between mobile and immobile regions. 
The expected solute breakthrough behavior is studied using numerical
random walk particle tracking simulations. This behavior is
analyzed by explicit analytical expressions for the asymptotic solute
breakthrough curves. We observe clear power-law tails of the solute
breakthrough for broad (power-law) distributions of particle transit
times (heterogeneous advection) and particle trapping times (MRMT
model). The combined model displays two distinct time regimes. An
intermediate regime, in which the solute breakthrough is dominated by
the particle transit times in the mobile zones, and a late time regime
that is governed by the distribution of particle trapping times in
immobile zones.   
\end{abstract}

\end{frontmatter}

\section{Introduction}
Solute transport in heterogeneous porous media displays behaviors that
cannot be captured by transport models based on an equivalent advection
dispersion equation (ADE) parameterized by (constant) effective
transport parameters. Such behaviors range from the non-linear
evolution of solute dispersion to power-law tails in solute
breakthrough curves~\cite[][]{BG1990, berkowitzcortis06}. The last
three decades have seen intense research to quantify these behaviors
in terms of effective transport models that can be obtained by
 moment equation approaches~\cite[][]{Neuman1987}, and projector
formalisms~\cite[][]{CHG1994}, for example, and include time and space fractional
ADEs~\cite[][]{Benson2000, Benson2004}, multirate mass transfer (MRMT)
models~\cite[][]{haggertyetal:1995WRR, Carrera1998}, as well as
continuous time random walks~\cite[][]{berkowitz1997, dentz:2004}, see
also the reviews in~\cite{berkowitzcortis06, Neuman_Tartakovsky2008,
  dentz:2011}.  

In this paper, we focus on the CTRW approach to modeling non-Fickian
solute transport in heterogeneous media. Classical random walks model
particle movements by using variable spatial steps which are taken within
constant time increments at equidistant
times~\cite[][]{delay_review2005, SalamonRW}. A CTRW, in contrast,
models particle movements in a heterogeneous medium effectively
as a random walk in which both space and time increments are
variable. The spatial transitions may
reflect the geometry of the underlying medium and flow heterogeneity,
while particle transition and waiting times reflect persistent particle
velocities over given transition distances, or particle retention due
to adsorption or diffusion into immobile
zones, for example~\cite[][]{berkowitz1997, leborgnedentz08-prl, DCa:2009,
 DentzBolster2010, kangetal11-prl}. The medium heterogeneity
is mapped into the probability distribution density (PDF) of characteristic particle transition
times. The evolution of the particle density, or, equivalently the
solute concentration is governed by a temporally non-local ADE whose
memory kernel is given in terms of the PDF of transition
times~\cite[][]{Berkowitz2002, dentz:2004}.

The MRMT approach is phenomenologically similar to the CTRW modeling
framework as it models the impact of medium heterogeneity on large
scale transport through a distribution of typical solute retention
times in immobile regions. In fact, it can be
shown~\cite[][]{schumer:2003, dentz:2003} that one model can under
certain conditions be mapped onto the other. The latter amounts
essentially to identifying the relation between the PDF of particle
transition times and particle retention
times in immobile regions~\cite[][]{Margolin:et:al:2003, BensonMRMT2009,
  DentzGouzeAWR2012}.  

As pointed out above, the CTRW model is a random walk
approach in that particle movements are governed by random walk equations
for the space and time coordinates. Therefore the solution of CTRW and
equivalent models is directly accessible to numerical solution through
random walk particle tracking simulations~\cite[][]{dentz:2004}. This
provides an avenue for the efficient simulation of transport in the
presence of mobile-immobile processes~\cite[][]{LeBorgneGouze2008, BensonMRMT2009,
DentzGouzeAWR2012}, for example, and for temporally non-local
transport in general~\cite[][]{Benson:2008}. 

Many formulations of the above models are for transport situations
under uniform mean flow. Thus, for the interpretation of tracer tests
under forced conditions they are only of limited applicability because
the non-stationarity of the underlying flow field is not accounted
for.  Haggerty et al.~\cite[][]{Haggerty:2001WRR} use a Eulerian radial MRMT
implementation to interpret radial single-well injection-withdrawal (SWIW)
tracer tests in fractured dolomite. Le Borgne and
Gouze~\cite[][]{LeBorgneGouze2008} used a CTRW based random walk
implementation of MRMT to model tracer breakthrough data from SWIW
tracer tests. Benson et al.~\cite[][]{Benson2004} developed a
fractional-order dispersion model in radial coordinates to model
tracer tests under forced conditions. 
A general issue when interpreting field tracer data is to decipher
the origin of the observed non-local transport behavior, which
may range from mobile-immobile diffusive mass transfer processes
to highly heterogeneous advective transport \cite[][]{Becker:2003wd,
  Kang2015}. In the latter case, non-Fickian transport may be
caused by a broad distribution of flow and transport velocities; the
distribution of particle transit times depends on the flow rate and
heterogeneity in the flow properties. In the former case, anomalous
transport features are due to mass transfer between mobiel and
immobile zones; particle transition times may depend on the retention
properties and geometries of the immobile regions.

Testing these different hypothesis requires non-local
transport models, that integrate both diffusive and advective
mass transfer processes in non-uniform flow conditions.

In this paper, we develop a general 
CTRW approach that allows
for the modeling of non-local solute transport under 
radial
conditions. The derivation from the space-time random walk equations
gives directly the particle tracking method for its numerical
solution. We present three non-local
CTRW based radial transport implementations, for the modeling of
heterogeneous advection, mobile-immobile mass transfer (MRMT), and the
combination of both. To this end we review in Section~\ref{RW} briefly the
random walk formulation of general radial advective-dispersive
transport. Section~\ref{Section:CTRW} then derives the general radial CTRW
framework and defines the specific CTRW models. The model breakthrough
curves then are analyzed in Section~\ref{Sec:BTC} using numerical random
walk simulations and explicit analytical expressions for the
asymptotic breakthrough behavior developed in~\ref{App:B}.
In particular, we discuss the expected differences in non-Fickian
transport behaviors induced by purely advective processes, purely diffusive processes,
and the combination of these processes. 





\section{Radial Random Walks\label{RW}}

The classical advection-dispersion equation (ADE) for the solute
concentration $c(r,t)$ in radial coordinates can be written as  
\begin{align}
\label{ADE}
  \frac{\partial c(r,t)}{\partial t} + \frac{1}{r}
  \frac{\partial}{\partial r} v(r) r c(r,t) - \frac{1}{r} \frac{\partial}{\partial r} r D(r)
\frac{\partial c}{\partial r}  = 0, 
\end{align}
where $v(r)$ and $D(r)$ are the radially dependent transport velocity
and dispersion coefficient; $r$ denotes the radial distance, $t$
denotes time. We set the constant porosity equal to one,
which is equivalent to rescaling time. 
The equivalent random walk particle tracking formulation is obtained
by rewriting ~\eqref{ADE} in mass conservative form. Therefore, we
define the conserved radial concentration as 
\begin{align}
\label{pmc}
p(r,t) = 2 \pi r c(r,t). 
\end{align}
Notice that $p(r,t)$ denotes the concentration per unit radial
distance.  
Inserting the latter into~\eqref{ADE} and rearranging terms we obtain
the radial Fokker-Planck equation
\begin{align}
\label{ADE:p:c}
\frac{\partial p(r,t)}{\partial t} + \frac{\partial}{\partial
  r} \left[v(r) + \frac{D(r)}{r} + D^\prime(r) \right]  p(r,t)   - \frac{\partial^2 }{\partial
  r^2} D(r) p(r,t) = 0,
\end{align}
where $D^\prime(r)$ denotes the derivative of $D(r)$ with respect to
$r$. The equivalent Langevin equation is given by \cite[][]{Risken:1996}
\begin{align}
\label{Langevin:r}
\frac{d r(t)}{d t} =v[r(t)] + \frac{D[r(t)]}{r(t)} + D^\prime[r(t)] +
\sqrt{2 D[r(t)]} \xi_r(t), 
\end{align}
where $\xi_r(t)$ is a Gaussian white noise of zero mean and the
correlation function $\langle \xi_r(t) \xi_r(t^\prime) \rangle =
\delta(t - t^\prime)$. Here and in the following, we employ the Ito interpretation~\cite[][]{Risken:1996}
of the Langevin equation~\eqref{Langevin:r}. The particle density is
given in terms of the radial trajectories as 
$p(r,t) = \langle \delta[r - r(t)] \rangle$, 
and by virtue of~\eqref{pmc}, we obtain for the concentration
distribution 
\begin{align}
\label{crt}
c(r,t) = \frac{\langle \delta[r - r(t)] \rangle}{2 \pi  r}. 
\end{align}

In the following, we will consider the case of~\cite[][]{Bear:1972}
\begin{align}
\label{choices}
v(r) = \frac{k_v}{r}, && D(r) = \frac{\alpha k_v}{r},
\end{align}
where $\alpha$ is dispersivity, and  $k_v = Q/(2 \pi)$ with $Q$ the
flow rate. Notice that more general radial dependences of flow
velocity and dispersion can be considered within the approaches
developed in the following. Here, we focus on the choice~\eqref{choices}.
With these definitions, the Langevin
equation~\eqref{Langevin:r} simplifies to 
\begin{align}
\label{Langevin:r:2}
\frac{d r(t)}{d t} = \frac{k_v}{r(t)} + \sqrt{\frac{2 \alpha
    k_v}{r(t)}} \xi_r(t). 
\end{align}
The temporally discretized version of the radial Langevin equation is
given by 
\begin{align}
\label{rw:discrete}
r_{n+1} = r_n + \frac{k_v \Delta t}{r_n} + \sqrt{\frac{2 \alpha
    k_v \Delta t}{r_n}} \xi_n, 
\end{align}
where $r_n = r(t_n)$, $t_n = n\Delta t$, and $\xi_n$ is a Gaussian
random variable with zero mean and unit variance. 

\section{Radial Continuous Time Random Walks\label{Section:CTRW}}
The radial random walk particle tracking formulations developed in the
following are based on the generalization of the radial random walk
process~\eqref{rw:discrete} in terms of the continuous time random
walk 
\begin{subequations}
\label{ccTRW}
\begin{align}
\label{cCTRW:a}
r_{n+1} &= r_n + \ell + \sqrt{2 \alpha \ell} \xi_n 
\\
\label{cCTRW:b}
t_{n+1} &= t_n + \tau_n(r),
\end{align}
\end{subequations}
where $\ell$ is a constant transition length, and $\tau(r)$ a radially
dependent, independently distributed random transition time
with the probability density function (PDF)
$\psi_\tau(\tau,r)$. Notice that the classical
formulation~\eqref{rw:discrete} is obtained
by setting $\tau_n(r) = \ell r_n/k_v$ in~\eqref{ccTRW}.  
The distribution of the spatial transition
lengths $\Delta r = \ell + \sqrt{2 \alpha \ell} \xi_n$ is denoted by
$\psi_r(\Delta r)$. The mean and mean square displacements are given
by $\langle \Delta r \rangle = \ell$ and $\langle \Delta r^2 \rangle =
2 \alpha \ell$, where we disregard contributions of order $\ell^2$. 
Notice that the transition length $\ell$ are chosen such that
$\ell \ll \alpha$.

A straightforward application of the general CTRW
framework~\cite[][]{berkowitzcortis06} gives for the radial particle density $p(r,t)$
the equation
\begin{align}
\label{p}
p(r,t) = \int\limits_0^t d t^\prime R(r,t^\prime) \int\limits_{t -
  t^\prime}^\infty d \tau \psi_\tau(\tau,r), 
\end{align}
where $R(r,t)$ denotes the probability per time that the particle has just
arrived at the radius $r$. This equation can be read as follows: The particle density at the position
$r$ at a time $t$ is given by the probability that a particle arrives
there at an earlier time $t^\prime$ times the probability that the
next transition takes longer than $t - t^\prime$. The density $R(r,t)$
satisfies the mass balance equation
\begin{align}
\label{R}
R(r,t) = \delta(r - r_0) \delta(t) + \int\limits_0^\infty d r^\prime
\int\limits_0^t dt^\prime R(r^\prime,t^\prime) \psi_\tau(t -
t^\prime,r^\prime) \psi_r(r - r^\prime),
\end{align}
where $r_0$ is the initial radial particle
position. Equations~\eqref{p} and~\eqref{R} can be combined into the
radial generalized Master equation
\begin{align}
\label{GME}
\frac{d p(r,t)}{d t} &= \int\limits_0^\infty d r^\prime \int\limits_0^t
d t^\prime \psi_r(r-r^\prime) \tau_k(r^\prime)^{-1} M(t -
t^\prime,r^\prime) p(r^\prime,t^\prime)  
\nonumber\\
&- \int\limits_0^t d t^\prime
\tau_k(r)^{-1}M(t - t^\prime,r)
p(r,t), 
\end{align}
where the memory kernel $M(t - t^\prime,r)$ is
defined in Laplace space by
\begin{align}
\label{M}
\hat M(\lambda,r) = \frac{\lambda \tau_k(r)  \hat
  \psi_\tau(\lambda,r)}{1 - \hat \psi_\tau(\lambda,r)}. 
\end{align}
The Laplace transform is defined in~\cite[][]{AS1972}. Laplace
transformed quantities in the following are marked by a hat, the
Laplace variable is denoted by $\lambda$. 

We consider here a $\psi_r(r)$ that is sharply peaked about its mean
value $\ell$. In this case, the generalized Master
equation~\eqref{GME} can be localized in space, and a Taylor
expansion of the integrand on the right side of~\eqref{GME} gives the
non-local radial Fokker-Planck equation
\begin{align}
\label{CTRW:p}
 \frac{\partial p(r,t)}{\partial t} + \int\limits_0^t dt^\prime\left(
 \frac{\partial }{\partial r} \frac{ k_v}{r} -   
 \frac{\partial^2}{\partial r^2} \frac{\alpha  k_v}{r}\right) M(t-t^\prime,r) p(r,t^\prime)=
 0.  
\end{align}
We disregard contributions of order $\ell^2$. Substituting the radial
concentration~\eqref{pmc} into this equation,
we obtain the non-local radial advection-dispersion equation 
\begin{align}
\label{CTRW:c}
 \frac{\partial c(r,t)}{\partial t} + \int\limits_0^t dt^\prime\left(
 \frac{ k_v}{r} \frac{\partial }{\partial r} -   \frac{\alpha  k_v}{r}
 \frac{\partial^2}{\partial r^2} \right) M(t-t^\prime,r) c(r,t^\prime)
= 0.  
\end{align}
Radial non-local partial differential equations such as~\eqref{CTRW:p}
and~\eqref{CTRW:c} can be solved subject to given initial and boundary
conditions, either numerically or semi--analytically using common
numerical schemes~\cite[][]{Benson2004,silva:2009} and analytical
methods. In this paper, we use random walk particle tracking to solve
for the transport behavior described by these equations. 

In the following sections, we present three different models of
increasing complexity for the time increment $\tau(r)$. The first one
emulates non-local transport due to advective heterogeneity, the
second model solves radial transport
under multirate mass transfer, the third model provides a random walk
approach combining heterogeneous advection and
multirate mass transfer into immobile zones in radial coordinates.

\subsection{Heterogeneous Advection \label{Sec:CTRW}}
We first consider a radial CTRW that represents non-Fickian
transport originating from broad distributions of advective transit times.
In radial flow conditions, the mean transit time increases linearly with the
radial distance. To represent this important property, we propose to scale
the random transit time $\tau(r)$ with the characteristic radial transition time
$\tau_k(r)$ as
%
\begin{align}
\label{CTRW}
\tau(r) = \tau_k(r) \eta, && \tau_k(r) = \frac{\ell r}{k_v}. 
\end{align}
The time scale $\tau_k(r)$ denotes the transition time over the space
increment $\ell \ll r$ under homogeneous flow conditions. The dimensionless
random increments $\eta$ are independent identically distributed according
to $\psi_\eta(\eta)$. They reflect the non-dimensional fluctuations of
the radial transport velocity due to spatial heterogeneity and are
related to the inverse radial flow velocity. Recently, Edery et
al.~\cite[][]{Edery2014} studied the relation of the distribution of
hydraulic conductivity and the transition time distribution under
uniform flow conditions. The distribution of inverse velocities, and
thus transition times, may be obtained from estimates of the hydraulic
conductivity distribution. Alternatively it may be modeled by a
parametric model~\cite[][]{Kang2015}, whose parameters are adjusted from the observed
breakthrough curves, which in turn may give insight into the flow
and medium heterogeneity.  

Thus, the transition times
$\tau(r) = \tau_k(r) \eta$ are distributed according to 
\begin{align}
\label{psi:tau:eta}
\psi_\tau(\tau,r) = \frac{1}{\tau_k(r)} \psi_\eta[\tau/\tau_k(r)]. 
\end{align}

The memory kernel $M(t,r)$ defined in the previous section can now be
written as
\begin{align}
\label{M_eta}
\hat M(r,\lambda) = \hat M_\eta[\lambda \tau_k(r)] \equiv \frac{\lambda
  \tau_k(r)  \hat \psi_\eta[\lambda \tau_k(r)]}{1 - \hat
  \psi_\eta[\lambda\tau_k(r)]}, 
\end{align}
where we used that the Laplace transform of~\eqref{psi:tau:eta} is
given by $\hat \psi(\lambda,r) = \psi_\eta[\lambda \tau_k(r)]$. Thus,
it follows that the memory kernel in real time has the scaling form 
$M(t,r) = \tau_k(r)^{-1} M_\eta[t/\tau_k(r)]$. The generalized
advection-dispersion equation~\eqref{CTRW:c} can then be written as
\begin{align}
\label{CTRW:c:ctrw}
 \frac{\partial c(r,t)}{\partial t} + \int\limits_0^t dt^\prime\left(
 \frac{ k_v}{r} \frac{\partial }{\partial r} -   \frac{\alpha  k_v}{r}
 \frac{\partial^2}{\partial r^2} \right)  \frac{M_\eta[(t - t^\prime)/\tau_k(r)]}{\tau_k(r)} c(r,t^\prime)
= 0.  
\end{align}
Notice the exponential distribution $\psi_\eta(\eta) =
\exp(-\eta)$ gives $M_\eta(\eta) = \delta(\eta)$ in~\eqref{M_eta} and thus the
generalized radial advection-dispersion equation~\eqref{CTRW:c:ctrw}
reduces to the radial advection-dispersion equation~\eqref{ADE} in a
homogeneous medium.  
\subsection{Multirate Mass Transfer\label{Sec:MRMT}}
The MRMT model considers solute transport under mass transfer between
a single mobile zone and a series of immobile
zones~\cite[][]{haggertyetal:1995WRR, HARV95, Carrera1998, dentz:2003,
  schumer:2003, silva:2009}. Here we derive a
radial random walk approach that simulates mass transfer between
mobile and immobile regions based on the radial CTRW~\eqref{ccTRW} and
the CTRW model presented in the previous section. 
An important difference with the advective CTRW model described in the
previous section is that the distribution of trapping times
does not depend on the radial position. 

Mass transfer between mobile and immobile
regions is modeled by a compound Poisson process for the transition
time $\tau(r)$ following the works of \cite{Margolin:et:al:2003},
\cite{BensonMRMT2009} and \cite{DentzGouzeAWR2012}. The particle
transition time $\tau(r)$ is split into a mobile time $\tau_m(r)$, which is the time
the particle needs to traverse the distance $\ell$ in the mobile
portion of the medium, and a series of immobile times $\tau_{im,i}$,
which measure the times the particles spent in the immobile portion
of the medium. The number of times $n_{im}$ a particle gets trapped
during a 
mobile transition of duration $\tau_{m}(r)$ is given by a Poisson
distribution characterized by the mean $\gamma \tau_m(r)$, where
$\gamma$ is the rate by which particles get trapped in immobile
zones,
\begin{align}
P_{im}[n,\tau_m(r)] = \frac{[\gamma \tau_m(r)]^n}{n!} \exp\left[-
  \gamma \tau_m(r) \right]. 
\end{align}
Thus, the time increment $\tau(r)$ in~\eqref{cCTRW:b}
associated to the spatial transition~\eqref{cCTRW:a} is given by 
\begin{align}
\label{compound}
\tau(r) = \tau_m(r) + \sum\limits_{i = 1}^{n_{im}} \tau_{im,i}.
\end{align}
The distribution of immobile times is denoted by $p_{im}(\tau)$. For
simplicity, we consider here the situation that initially all
particles are mobile. The mobile times are given by $\tau_m(r) = \tau_k(r) \eta$ as
in~\eqref{CTRW}, with an exponentially distributed $\eta$. Notice
that, as pointed out at the end of the previous section, an
exponential $\eta$ models transport in a homogeneous medium. This
means, $\tau_m(r)$ is exponentially distributed with mean
$\tau_k(r)$. Thus the transition time PDF can be expressed in terms of
its Laplace transform as (see~\ref{App:A})
\begin{align}
\label{psi:c:exp}
\hat \psi_\tau(\lambda,r) = \frac{1}{1 + \lambda \tau_k(r) + \gamma \tau_k(r)
[1 - \hat p_{im}(\lambda)]}. 
\end{align}
Inserting this expression into~\eqref{M} for the Laplace transform
$\hat M(\lambda,r)$ of the memory kernel gives the compact expression 
\begin{align}
\hat M(\lambda,r) \equiv \hat M(\lambda) = \frac{1}{1 + \gamma
  \lambda^{-1} [1 - \hat p_{im}(\lambda)]}. 
\end{align}
Note that the memory kernel is independent on the radial distance,
which reflects the difference between mobile particle transitions,
which depends on the local flow velocity, and particle retention in
immobile zones, which depends on the distribution of diffusion times,
for example. Using this expression in the Laplace transform of~\eqref{CTRW:c}
gives for the radial concentration
\begin{align}
\label{CTRW:MRMT}
\lambda \hat c(r,\lambda) + \left(
 \frac{ k_v}{r} \frac{\partial }{\partial r} -   \frac{\alpha  k_v}{r}
 \frac{\partial^2}{\partial r^2}\right) \frac{\hat c(r,\lambda)}{1 +
 \gamma \lambda^{-1}[1 - \hat p_{im}(\lambda)]} = c(r, t = 0).   
\end{align}
We define now the mobile solute concentration by its Laplace
transforms as 
\begin{align}
\hat c_m(r,\lambda) = \frac{\hat c(r,\lambda)}{1 + \gamma
  \lambda^{-1}[1 - \hat p_{im}(\lambda)]}. 
\end{align}
Thus, we obtain from~\eqref{CTRW:MRMT} for $\hat c_m(r,\lambda)$
\begin{align}
&\lambda \hat c_m(r,\lambda) +  \lambda \left\{\gamma \lambda^{-1}[1 - \hat
p_{im}(\lambda)] \hat c_m(r,\lambda)\right\} 
\nonumber\\
& \qquad \qquad + \left(
 \frac{ k_v}{r} \frac{\partial }{\partial r} -   \frac{\alpha  k_v}{r}
 \frac{\partial^2}{\partial r^2}\right) \hat c(r,\lambda) = c_m(r, t = 0).   
\label{MRMT:2}
\end{align}
Note that we assume for simplicity that initially all particles are
mobile. An initial presence of particles in the immobile zones would
give rise to a source term in~\eqref{MRMT:2}~\cite[][]{Tecklenburg2013}. 
We furthermore define the density of immobile particles by the expression
in curly brackets on the left side of~\eqref{MRMT:2}. It reads in time space as 
\begin{align}
c_{im}(r,t) = \gamma \int\limits_0^t dt^\prime c_{m}(r,t^\prime)
\int\limits_{t - t^\prime}^\infty d\tau p_{im}(\tau). 
\end{align}
The right hand side expresses the density of immobile particles by the
probability per time that mobile particles get trapped at a given time
$t^\prime$, $\gamma c_m(r,t^\prime)$, times the probability that the
residence time in the immobile region is larger than $t -
t^\prime$. Thus, at asymptotically long times, the ratio of the
time averaged immobile and mobile concentrations is given by  
\begin{align}
\lim_{t \to \infty} \frac{\overline c_{im}(r,t)}{\overline c_{m}(r,t)} = \lim_{\lambda \to 0} \gamma \lambda^{-1} [1 -
p_{im}^\ast(\lambda)] = \gamma \langle \tau_{im} \rangle, 
\end{align}
where $\langle \tau_{im} \rangle$ is the mean immobile time and the
overline denotes the time average $\overline c(r,t) = t^{-1} \int_0^t
dt^\prime c(r,t^\prime)$. 

We now define the memory function $\varphi(t)$ as 
\begin{align}
\label{phi}
\varphi(t) = \frac{1}{\langle \tau_{im} \rangle} \int
\limits_{t}^\infty d \tau p_{im}(\tau). 
\end{align}
Thus, the governing equation~\eqref{MRMT:2} of the mobile solute
concentration can be written in time space as
\begin{align}
\label{MRMT:3}
\frac{\partial c_m(r,t)}{\partial t} +  \beta \int\limits_0^t d
t^\prime \varphi(t - t^\prime) c_m(r,t^\prime) 
+ \left(\frac{ k_v}{r} \frac{\partial }{\partial r} -   \frac{\alpha  k_v}{r}
 \frac{\partial^2}{\partial r^2}\right) c(r,t) = 0. 
\end{align}
where we defined $\beta = \gamma \langle \tau_{im}
\rangle$. Equation~\eqref{MRMT:3} describes transport under multirate
mass transfer in radial flow~\cite[][]{Haggerty:2001WRR,
  LeBorgneGouze2008}. 
The memory function $\varphi(t)$ encodes the mass transfer
mechanism~\cite[][]{haggertyetal:1995WRR, HARV95, Carrera1998,
  dentz:2003, DGC2011} between the mobile and immobile regions. 
For linear first-order mass exchange it reflects the distribution of transfer
rates between mobile and immobile
regions~\cite[][]{haggertyetal:1995WRR}. For diffusive mass
transfer, it encodes the geometries and the characteristic diffusion
scales of the immobile regions~\cite[][]{Noetinger:2000, GMDC2008}. For transport
through highly heterogeneous porous and fractured media, the memory
function may be related semi-analytically or empirically to the medium
heterogeneity~\cite[][]{Willmann2008, Zhang2014}.  

The memory
function is here defined in terms of the distribution of residence
times in the immobile regions given by~\eqref{phi}. Reversely, the
distribution of residence times $p_{im}(\tau)$ can be obtained for a
given memory function $\varphi(t)$ according to 
\begin{align}
\label{pim}
p_{im}(\tau) = - \langle \tau_{im} \rangle \frac{d \varphi(\tau)}{d \tau}. 
\end{align}
Expressions~\eqref{phi} and~\eqref{pim} establish the relation between
the distribution of residence times and the memory function. 
\subsection{Combination of Heterogeneous Advection and Multirate
  Mass Transfer\label{Sec:CTRWMRMT}}
We consider now the random walk implementation of the combination of
CTRW, which accounts for heterogeneous transport in the mobile zone,
and multirate mass transfer between the mobile and immobile regions. 
Using the approach presented in the previous section, the mobile
transition time $\tau_m(r)$ in~\eqref{compound} is now given by the
general relationship $\tau_m(r) = \tau_k(r) \eta$. The random variable
$\eta$ is distributed according to a general $\psi_\eta(\eta)$ as
outlined in Section~\ref{Sec:CTRW}. Accordingly, the PDF
for the mobile transitions is given by~\eqref{psi:tau:eta} and reads
as 
\begin{align}
\label{psi_m_eta}
\psi_m(\tau,r) = \frac{1}{\tau_k(r)} \psi_\eta[\tau/\tau_k(r)]. 
\end{align}
Thus, the transition time PDF for the general compound~\eqref{compound} process reads
in terms of its Laplace transform as (see~\ref{App:A})
\begin{align}
\label{psi:compound}
\hat \psi_\tau(\lambda,r) = \hat \psi_m\left(\lambda + \gamma [1 -
  \hat p_{im}(\lambda)],r \right).
\end{align}
Specifically, by using the Laplace transform of~\eqref{psi_m_eta},
$\hat \psi_m(\lambda,t) = \hat \psi_\eta[\lambda \tau_k(t)]$, the
Laplace transform of $\psi_\tau(\tau,r)$ can be written as 
\begin{align}
\label{psi:het:comb}
\hat \psi_\tau(\lambda,r) = \hat \psi_\eta\left(\lambda \tau_k(r) +
   \gamma \tau_k(r) [1 - \hat p_{im}(\lambda)],r \right).
\end{align}
Note that unlike for the mobile immobile model discussed in the
previous section, the transition time PDF~\label{psi:het:mrmt} renders
the memory kernel~\eqref{M} dependent on the radial distance. 
The total trapping time during a mobile transition is related to the
number of trapping events, whose mean is given by the mobile
transition time times the trapping rate. While the individual trapping
times are independent on advective heterogeneity, their collective is
related to advective heterogeneity through the number of trapping
event. This interrelation is reflected in the radial dependence
of~\eqref{psi:het:comb}.  

For times $t \ll \gamma^{-1}$, or equivalently $\lambda \gg \gamma$,
the transition time PDF~\eqref{psi:het:comb} can be approximated by
the one for purely advective heterogeneity. Thus for $t \ll
\gamma^{-1}$, transport is dominated by advective heterogeneity, which
is evident because at $t \ll \gamma^{-1}$ the number of trapping
events is very small. If the PDF of dimensionless mobile transition times
$\psi_\eta(\eta)$ has the finite mean transition time $\langle \eta
\rangle < \infty$,  the transition time PDF~\eqref{psi:het:comb} can be
approximated for small arguments as 
\begin{align}
\label{psi:approx:8}
\hat \psi_\tau(\lambda,r) = 1 - \tau_k(r) \langle \eta \rangle \lambda
\left\{1 + \lambda^{-1} \gamma [1 - \hat p_{im}(\lambda)]\right\}.
\end{align}
Notice that this is valid as long as $\lambda \tau_k(r) + \gamma
\tau_k(r) [1 - \hat p_{im}(\lambda)] \ll 1$. 
Inserting~\eqref{psi:approx:8} into~\eqref{M} gives in leading order
for the memory kernel
\begin{align}
\hat M(\lambda,r) \equiv \hat M(\lambda) = \langle \eta \rangle^{-1} 
\frac{1}{1 + \gamma \lambda^{-1} [1 - \hat p_{im}(\lambda)]}.
\end{align}
Thus, in this limit, the model behaves as the MRMT model introduced in
the previous section.  These asymptotic behaviors of the different
models are discussed in more detail in the next section, which studies
solute breakthrough in the presented CTRW models. 
\section{Breakthrough Curves\label{Sec:BTC}}
We discuss here the behavior of solute breakthrough curves 
in the radial non-local models developed in the previous section
for an instantaneous pulse tracer injection at radius $r = r_0$.
Notice that the response to a pulse has a fundamental character
because responses to other injection conditions can be obtained by
superposition of pulse responses. Furthermore, in field tracer
experiments the duration of the tracer pulse is typically much shorter
than the duration of the experiment and may be approximated as
instantaneous~\cite[][]{Becker:2003wd, Benson2004, Kang2015}.  
 
We use numerical random walk particle tracking simulations to solve for the
breakthrough behavior in each of the models and derive explicit
analytical expressions for their asymptotic behaviors. The numerical
simulations are based on the recursion relations~\eqref{ccTRW} for the
particle positions and times. Specifically, the initial position at $n
= 0$ random walk steps is set to $r_0 = \ell$ and the initial time is
set to $t_0 = 0$. 

The solute breakthrough curve is identical to the distribution
$f(t,r)$ of first passage times of solute particles at the radius
$r$. The first passage time $\tau_a(r)$ is defined by 
\begin{align}
\tau_a(r) = t_{n_r}
\end{align}
where $n_r = \min(n|r_n \geq r)$ denotes the minimum number of steps
needed to pass the radius $r$. The first passage time PDF then 
is given by 
\begin{align}
f(t,r) = \sum\limits_{n=0}^\infty f_0(n,r) p(t,n), 
\end{align}
where $f_0(n,r)$ denotes the distribution of the minimum number of steps
needed to exceed $r$, and $p(t,n)$ the distribution of times after $n$
random walk steps in~\eqref{cCTRW:b}. 
\subsection{Heterogeneous Advection}
The CTRW approach representing heterogeneous advection models the
transition time by the time increment~\eqref{CTRW}, which is
determined by the dimensionless increment $\eta$. In the following, we
consider for $\psi_\eta(\eta)$ the Pareto distribution
\begin{align}
\label{eta:pareto}
\psi_\eta(\eta) = \beta
\eta^{-1 - \beta}, && \eta \geq 1, 
\end{align}
which models a broad distribution of transport velocities. 
This type of pure power-law behavior may be observed in an
intermediate time regime. Asymptotically one would expect that the
transition time PDF is truncated on a scale corresponding to a largest
heterogeneity scale, for example, see also the discussion
in~\cite[][]{dentz:2004}. 

For a power-law distribution $\psi_\eta(\eta)$ of dimensionless
transition times, we derive in~\ref{App:B:1} the following scaling
forms of the breakthrough curves for times larger than $\tau_k(r)$   
\begin{align}
\label{p:adv1:rt}
f(t,r) &= \frac{1}{\theta(r)} f_{01}\left[\frac{t}{\theta(r)}\right], & 0 &< \beta <
1\\
\label{p:adv2:rt}
f(t,r) &= \frac{1}{\theta(r)} f_{02}\left[\frac{t - \langle
    \tau_{a}(r)\rangle}{\theta(r)} \right] , & 1 &< \beta < 2, 
\end{align}
where we defined
\begin{align}
\label{thetatau}
\theta(r) = \left[\sum\limits_{i=1}^{r/\ell} \tau_k(r_i)^\beta
\right]^{1/\beta}, && \langle \tau_a(r) \rangle = \frac{1}{\beta - 1} \sum\limits_{i=1}^{r/\ell}
\tau_k(r_i). 
\end{align}
We furthermore obtain for the asymptotic behavior of the
scaling functions $f_{01}(x)$ and $f_{02}(x)$ for $x \gg 1$ the power-law decay
$\sim x^{-1-\beta}$ such that we obtain for $t \gg \theta(r)$ the
following behavior for the breakthrough curves
\begin{align}
\label{fhetadv}
f(t,r) \sim \frac{1}{\theta(r)}
\left[\frac{t}{\theta(r)}\right]^{-1-\beta}. 
\end{align}

For $0 < \beta < 1$, mean and variance of the arrival time do not
exist, for $1 < \beta < 2$, the mean exists and is given by  
$\langle \tau_a(r) \rangle$, while the variance diverges. The mean
arrival time depends on distribution of transition times through $\beta -1$ in the
denominator. The second term in the expression for $\langle \tau_a(r) \rangle$ is the mean
arrival time for a homogeneous model characterized by an exponential
PDF of dimensionless transition times $\eta$.
For $0 < \beta < 1$, the characteristic time $\theta(r)$ scales
peak width and position, as can be deduced from the
scaling form~\eqref{p:adv1:rt}. For $1 < \beta < 2$, it is a measure
for the peak width, as indicated by~\eqref{p:adv2:rt}. 
Using $\tau_k(r) = \ell  r /k_v $, we obtain for $\theta(r)$ and $\langle
\tau_a(r) \rangle$
%
\begin{align}
\label{theta}
\theta(r) \approx \frac{\ell^2}{k_v} \left(\frac{1}{1+\beta}
\right)^{1/\beta} \left(\frac{r}{\ell} \right)^{1+1/\beta}, &&  
\langle \tau_a(r) \rangle \approx \frac{1}{\beta - 1} \frac{r^2}{2 k_v}. 
\end{align}
Note that the late time scaling~\eqref{fhetadv} is the same as for a uniform flow
scenario that is characterized by a power-law distribution of
transition times~\cite[][]{dentz:2004}. The radial geometry is
reflected in the dependence of the characteristic time
scales~\eqref{thetatau} on $r$. 



\begin{figure}[t]
\includegraphics[width=.45\textwidth]{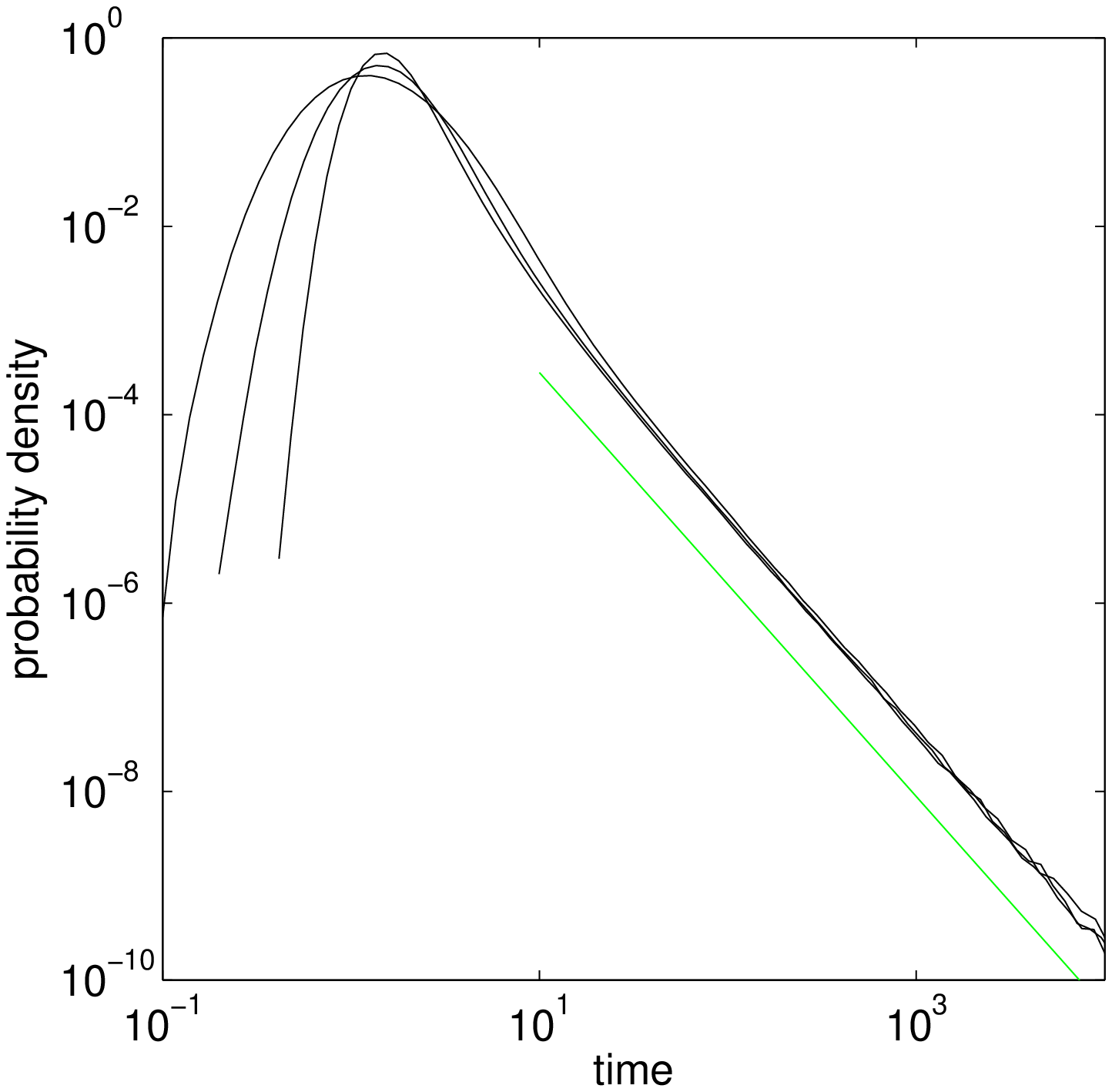}
\includegraphics[width=.45\textwidth]{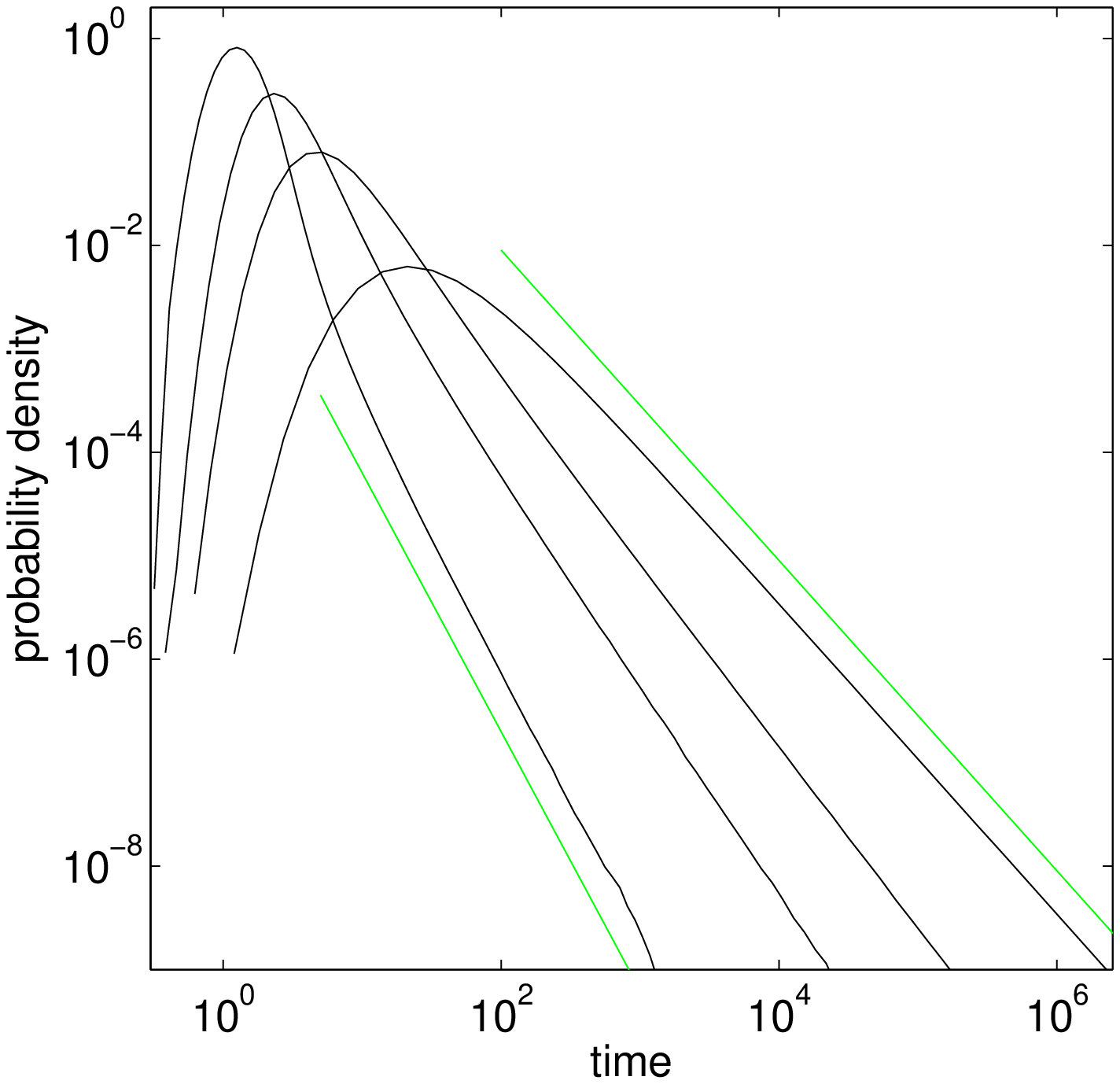}
\caption{Breakthrough curves for the CTRW model~\eqref{CTRW} with (a) $\alpha = 10^{-2}, 4 \times
  10^{-2}, 10^{-1}$, $\beta = 5/4$, and (b) (from left to right) $\beta = 3/2, 1, 3/4, 1/2$, and
  $\alpha = 2 \times 10^{-2}$. The peak width in (a) increases with
  increasing $\alpha$. The green lines show the power-law $\sim t^{-1
    - \beta}$ behaviors for (a) $\beta = 5/4$
  and (b) $\beta = 1/2$ and $\beta = 3/2$. 
  \label{fig1}}
\end{figure}
Figure~\ref{fig1} shows solute breakthrough curves for different
values of dispersivity $\alpha$ and exponents $\beta$ of the Pareto
distribution~\eqref{eta:pareto}. 
The width of the peak of the breakthrough curves increases with
increasing $\alpha$, as shown in Figure~\ref{fig1}a. The peak arrival
time remains unaffected by $\alpha$. Figure~\ref{fig1} shows the
behavior of the breakthrough curves
for different exponents $\beta$. With increasing $\beta$, the peak
arrival time and the width of the breakthrough peak both decrease as
indicated by~\eqref{theta}. For
times $t \gg \tau_k(r)$ we clearly observe the power-law
behavior~\eqref{fhetadv}. As outlined above, the late time power-law
behavior is the same as the one observed for uniform form. The scaling
of peak arrival and width, however, are impacted on by both the radial
geometry and the transition time PDF, as given in~\eqref{thetatau}.


\subsection{Multirate Mass Transfer}

The CTRW model for multirate mobile-immobile mass exchange is
characterized by the transition times~\eqref{compound}. The mobile
times are given by $\tau_m(r) = \tau_k(r) \eta$, where $\eta$ here is
distributed according to the exponential PDF
\begin{align}
\label{psi_exp}
\psi_\eta(\eta) = \exp(-\eta). 
\end{align}
The distribution of immobile times $p_{im}(\tau)$ is given by the
Pareto distribution 
\begin{align}
\label{pim:ex}
p_{im}(\tau) = \frac{\delta}{\tau_c} \left(\frac{\tau}{\tau_c}\right)^{-1 -
\delta}, && \tau \geq \tau_c.
\end{align}
Notice that the memory function $\varphi(t)$ for diffusive mass
transfer between a mobile and homogeneous immobile zones behaves as
$\varphi(t) \sim t^{-1/2}$~\cite[][]{haggertyetal:1995WRR,Carrera1998}. Thus, we obtain
from~\eqref{pim} that the corresponding PDF of immobile times behaves
as $p_{im}(\tau) \sim t^{-3/2}$, which corresponds to $\delta = 1/2$
in~\eqref{pim:ex}. For heterogeneous immobile zones, one obtains in
general different behaviors for the memory function
$\varphi(t)$~\cite[][]{GMDC2008} and therefore for the PDF of immobile
times. The asymptotic behavior of the breakthrough curves for times $t
\gg \tau_c$ is derived in~\ref{App:B:2}. It is given by 
\begin{align}
\label{fMRMT}
f(t,r) \sim \frac{1}{\theta_{c}(r)} \left[\frac{t}{\theta_{c}(r)}\right]^{-1 - \delta}, &&
\theta_{c}(r) = \tau_c \left[\gamma  \sum\limits_{i = 1}^{r/\ell} \tau_k(i\ell)
\right]^{1/\delta}. 
\end{align}
%
Recall that $\gamma$ is the rate for trapping in the immobile zones. 
The characteristic time $\theta_c(r)$ is a measure for peak
width and the time for the onset of the power-law tail behavior
$f(t,r) \sim t^{-1 - \delta}$. Using the explicit $\tau_k(r) = \ell r
/ k_v$ gives for $\theta_c(r)$
\begin{align}
\theta_c(r) \approx \tau_c \left(\frac{\gamma r^2}{2 k_v} \right)^{1/\delta}
\end{align}
%

\begin{figure}[t]
\includegraphics[width=.45\textwidth]{Figure2a}
\includegraphics[width=.45\textwidth]{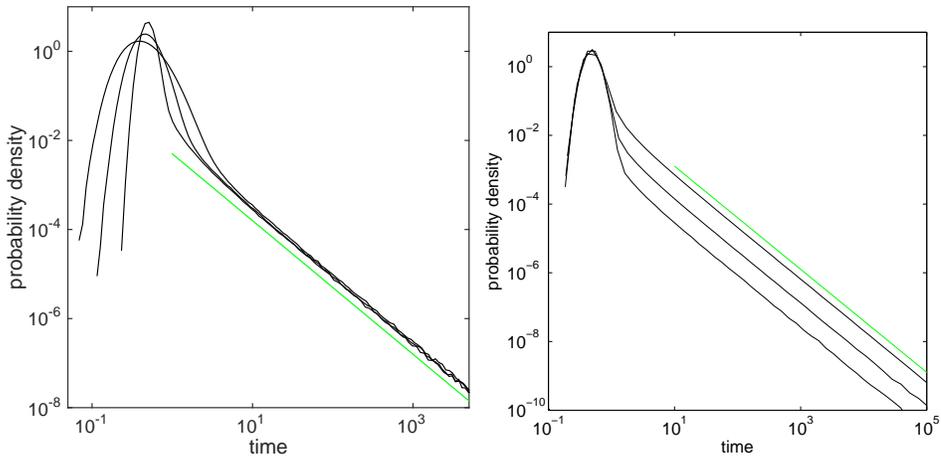}
\caption{Breakthrough curves for the MRMT model~\eqref{compound} with (a) $\alpha = 10^{-2}, 4
  \times 10^{-2}, 10^{-1}$, $\delta = 1/2$, $\tau_c = 10^{-1}$ and
  $\gamma = 10^{-1}$, and (b) $\alpha = 2 \times 10^{-2}$, $\delta =
  1/2$, $\tau_c = 10^{-1}$ and $\gamma = 10^{-2}, 5 \times 10^{-2},
  2.5 \times 10^{-1}$. The green lines show the power-law $\sim t^{-1
    - \delta}$ behavior for $\delta = 1/2$. 
\label{fig2}}
\end{figure}
Figure~\ref{fig2} shows the behavior of the breakthrough curves for
$\delta = 1/2$ and varying dispersivity $\alpha$ and trapping rate
$\gamma$. As for the case of heterogeneous advection, increase in
dispersivity $\alpha$ leads to an increase of the width of the
breakthrough peak, as illustrated in Figure~\ref{fig2}a. The peak
arrival time remains unchanged. Figure~\ref{fig2}b shows solute
breakthrough curves for varying trapping rates $\gamma$. 
As $\gamma$ increases, the weight in the power-law tail increases as
well as the peak width and therefore also the time of the onset of the power-law
behavior. Note that increasing the trapping rate increases the
proportion of trapped particles at any time, which explains the
increased weight in the tail of the breakthrough curve. At the same
time, increasing $\gamma$ decreases the time that particles spend mobile,
and thus particularly the average time until which particles get
trapped for the first time. Thus, as particles notice the presence of
immobile zones earlier, the peak width increases, and the breakthrough
curve breaks off earlier from the behavior without immobile zones. 

 
\subsection{Heterogeneous Advection with Multirate Mass Transfer}
The CTRW model combining heterogeneous advection in the mobile zone
with mass exchange between mobile and immobile regions is based on the
transition times~\eqref{compound}. The mobile time increment is again
given by $\tau_{m}(r) = \tau_k(r) \eta$, where $\eta$ now is modeled by
the Pareto distribution~\eqref{eta:pareto} characterized by the
exponent $\beta$. The PDF of immobile times is again given by the
Pareto distribution~\eqref{pim:ex} characterized by the exponent
$\delta$. We focus here on situations, for which $\beta >
\delta$. Notice that the exponents $\beta$ and $\delta$ encode the
width of the distributions of characteristic advection scales and
trapping time scales, respectively. 
The width of characteristic
trapping time scales is typically larger than the one of
characteristic advection time scales in the mobile zone, and thus we
set $\beta > \delta$.  

\ref{App:B:3} develops the asymptotic breakthrough behavior
for this model. One can distinguish two time regimes with distinct
temporal behavior. An intermediate time regime is set by the advection
scale $\tau_k(r)$ and the immobile time scale $\hat \tau_{im} = \tau_c
(\alpha_{im} \gamma\tau_c)^{\frac{1}{\delta -1}}$ for $\tau_k(r) \ll \hat
\tau_{im}$. We consider the case $0 < \delta < 1$ such that this
condition can only be achieved for $\gamma \tau_c \ll 1$, which
implies a low trapping rate. Thus, in this time regime, most particles
have not yet encountered a trapping event and thus, the breakthrough
curve behaves as in the case of heterogeneous advection given
by~\eqref{fhetadv}, $f(t,r) \sim t^{-1-\beta}$. 

In the long time regime $t \gg \hat \tau_{im}$, we need to distinguish
between the cases $0 < \beta < 1$ and $1 < \beta < 2$. For $0 < \beta
< 1$, the long time breakthrough behavior is given by
\begin{align}
f(t,r) \sim t^{-1-\beta\delta}.
\end{align}
Note that $0 < \beta < 1$ indicates quite strong tailing in the mobile
zone. Thus both the exponents characteristic for the mobile and
immobile zones determine the long time breakthrough behavior. 
For $1 < \beta < 2$ the long time breakthrough behavior is dominated
by particle retention due to trapping in the immobile regions. The
breakthrough curve behaves asymptotically as given by~\eqref{fMRMT}.  

\begin{figure}[t]
\includegraphics[width=.45\textwidth]{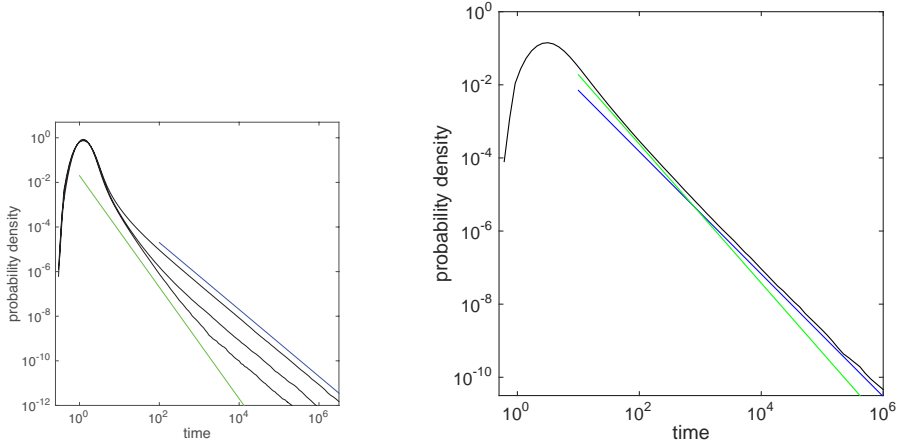}
\includegraphics[width=.45\textwidth]{Figure3b}
\caption{Breakthrough curves for the combined model~\eqref{compound}
  with (a) $\alpha = 2 \times 10^{-2}$, $\beta = 3/2$, $\delta = 1/2$,
  $\tau_c = 1$ and (from top to bottom) $\gamma = 10^{-2}, 10^{-3},
  10^{-4}$, (b) $\alpha = 2 \times 10^{-2}$, $\beta =
  0.9$, $\delta = 3/4$, $\tau_c = 1$ and $\gamma = 5 \times 10^{-2}$.
The green lines indicates the power-law $\sim t^{-1-\beta}$ for (a)
$\beta = 3/2$ and (b) $\beta = 0.9$. The blue line in (a) indicates
the power-law $\sim t^{-1-\delta}$
  for $\delta = 1/2$. The
  blue line in (b) indicates the
  power-law $\sim t^{-1-\beta \delta}$ for $\beta = 0.9$ and $\delta =
  3/4$. 
\label{fig3}}
\end{figure}

These behaviors are illustrated in
Figure~\ref{fig3}. Figure~\ref{fig3}a shows the behaviors in the
intermediate and asymptotic long-time regimes for $\beta = 3/2$ and
$\delta = 1/2$. We see a relatively short intermediate regime (which
increases as the trapping rate $\gamma$ decreases)
dominated by the advective heterogeneity and a the long time regime
dominated by particle retention in the immobile regions. 
A very similar behavior has been observed in the field from
push pull tracer tests \citep{LeBorgneGouze2008} pointing
to a combined effect of advective and diffusive processes on non-Fickian transport.
Figure~\ref{fig3}b
illustrates the breakthrough curve for $\beta = 0.9$ and $\delta =
0.75$. The intermediate time regime again is dominated by the
advective heterogeneity while the asymptotic tailing behavior is
determined by both the advective and trapping exponents. 

\section{Summary and Conclusions}

The CTRW approach provides a versatile framework for the modeling of
non-Fickian solute transport in heterogeneous media. In this paper we
develop a general CTRW approach for transport under radial flow
conditions starting from the random walk equations for the radial and
temporal particle coordinates. In contrast to CTRW formulations under
uniform flow conditions here the random transition times form a
non-stationary stochastic process, which reflects the radial
dependence, i.e., non-stationarity, of the flow conditions. Thus, the
evolution of the solute concentration is governed by a non-local ADE
that is characterized by a radially dependent memory kernel. 
Within this general framework, we present CTRW models that
may account for the impact of advective heterogeneity on large scale
transport, that implement multirate mass transfer between a mobile
immobile regions, and that combine non-local mobile transport due to
heterogeneous advection and mass
transfer into immobile zones. The three models are characterized by
the specific forms of the stochastic process of particle times. 

For heterogeneous advection, the transition time reflects the inverse
(heterogeneous) flow velocity, which in average depends on the total
flow rate and on the radial distance. This is accounted for by a
transition time that is modeled as the product of a non-dimensional
time increment and an advective transition time that is proportional
to the radial distance and the inverse flow rate. Thus, transition
times decrease with increasing flow rate as expected for purely
advective heterogeneity. Flow heterogeneity is accounted for by the
distribution of the dimensionless time increment~\cite[][]{Kang2015}.  
The memory kernel of the non-local ADE for solute concentration
depends here on the radial distance. 

Radial transport under multirate mass transfer between mobile and
immobile regions is implemented by separating the transition time into
the time the particle is mobile and the sum of immobile times, which
is represented by a compound Poisson
process~\cite[][]{Margolin:et:al:2003, BensonMRMT2009} whose mean is
given by the trapping rate times the local mobile trapping time. The mobile
time is modeled by an exponentially distributed dimensionless time
increment, which models Fickian mobile transport. The mean number of
trapping events during a mobile transition is given by the trapping
rate and the mean mobile time. The memory kernel
here is independent of the radial distance, which reflects the fact
that the individual trapping events do not depend on mobile
advection, which is consistent with radial MRMT formulation that
assume homogeneous advection in the mobile
zone~\cite[][]{Haggerty:2001WRR}. 

Finally, we propose a CTRW model that combines heterogeneous advection
in the mobile continuum with multirate mass transfer into immobile
continua. The mobile transition time now is modeled by a
non-exponential dimensionless time increment. The immobile time
increment is again a compound Poisson process. While the individual
trapping events are independent on the heterogeneous advection, the
collective of trapping times depends on the radial distance
through the mean number of trapping events. Unlike for the case of a
homogeneous mobile advection, heterogeneity leads to an interrelation
of the PDFs of mobile transition times and retention times in immobile
regions, and thus to a radial dependence of the memory kernel.   

The derivation of the general framework from the random walk equations
for the particle coordinates in space and time provides directly the
random walk particle tracking methods for the numerical
solution of radial MRMT models, and temporally non-local radial
transport formulations in general. Notice that the developed non-local
transport models assume that dispersion is proportional to the flow
velocity, as well as a constant injection or withdrawal rate of the
flow at the well. It is straightforward to relax the first assumption
and introduce a more general velocity dependence of dispersion in the
developed models. The assumption of a constant flow rate may be
relaxed to account for scenarios encountered in single well injection
withdrawal tests, i.e., piecewise constant flow rates.  

The solute breakthrough behaviors for the different non-local radial
transport models are studied in detail by using random walk particle
tracking simulations and explicit analytical expressions for the
asymptotic time behaviors of the solute breakthrough curves. A broad
distribution of advective transition times is modeled by a Pareto
distributions of dimensionless times. The distribution of immobile
times is also given by a Pareto distribution, which mimics a
broad distribution of diffusion or in general retention times
in the immobile zones. We find distinct power-law tail behavior in all
cases, which are similar to the ones that can be observed for uniform
flow. However, the non-stationarity of the random time increment leads
to a non-trivial radial dependence of the breakthrough curves. 
The model combining heterogeneous advection and MRMT can display an
intermediate time regime in which heterogeneous advection dominates,
before the onset of trapping, and a late time regime that is governed
by the retention properties in the immobile regions. The different
behaviors in the two time-regimes allow for the discrimination and
identification of heterogeneous advection and mobile-immobile mass
transfer as drivers of anomalous transport. The CTRW
model accounting for advective heterogeneity is controlled by the flow
rate, while the CTRW model for MRMT is controlled by the trapping rate
and the properties of the immobile zone. 


In conclusion, this paper provides a radial CTRW framework that allows
for the systematic interpretation of tracer data from radial forced gradient
test, through the implementation of different mechanisms
that lead to non-local radial transport. 


\bigskip 

\noindent
{\bf Acknowledgements:} MD acknowledges the support of the European
Research Council (ERC) through the project MHetScale (contract number
617511). We thank Dr. Antoine Aubenau and two anonymous reviewers for
their comments. 

\appendix
\section{Derivation of the Transition Time Distribution for the
  Compound Time Process\label{App:A}}
We derive here the probability density function (PDF) $\psi(t,r)$ for the
general compound process~\eqref{compound} of the time increment
$\tau(r)$. The density of the mobile time $\tau_m(r) = \tau_k(r) \eta$ is denoted by 
$\psi_m(\tau,r)$, and can be expressed in terms of the  PDF
$\psi_\eta(\eta)$ as 
\begin{align}
\psi_m(\tau,r) = \frac{1}{\tau_k(r)} \psi_\eta[\tau/\tau_k(r)]. 
\end{align}

The density of the compound process $\tau(r)$ can be written in
general as
\begin{align}
\psi_\tau(\tau,r) = \left\langle \delta[\tau - \tau(r)] \right\rangle,  
\end{align}
where the angular brackets denote the ensemble average over all
particles. Inserting~\eqref{compound} and performing the average
over the $\tau_m(r)$ gives 
\begin{align}
\psi_\tau(\tau,r) = \int\limits_0^\infty d \tau_m \psi_m(\tau_m,r)
\left\langle \delta\left(\tau - \tau_m - \sum_{i = 1}^{n_{im}} \tau_{im,i} \right) \right\rangle
\end{align}
Performing the Laplace transform in $\tau$ and performing the average
of the $\tau_{im,i}$ gives 
\begin{align}
\hat \psi_\tau(\lambda,r) = \int\limits_0^\infty d \tau_m
\psi_m(\tau_m) \exp(-\lambda \tau_m)
\left\langle \hat p_{im}(\lambda)^{n_{im}} \right\rangle.
\end{align}
Performing the average of the Poisson variable $n_{im}$ results in
\begin{align}
\hat \psi_\tau(\lambda,r) = \int\limits_0^\infty d \tau_m \psi_m(\tau_m)
\sum_{n = 0}^\infty \frac{\hat p_{im}(\lambda)^{n} (\gamma
  \tau_m)^n}{n!} \exp[- (\lambda + \gamma) \tau_m].   
\end{align}
The infinite series can be summed up to an exponential such that 
\begin{align}
\hat \psi_\tau(\lambda,r) = \int\limits_0^\infty d \tau_m \psi_m(\tau_m)
\exp\left(-\left\{\lambda + \gamma [1 - \hat p_{im}(\lambda)] \right\}
\tau_m\right).
\end{align}
The integration over $\tau_m$ can be performed explicitly by noting that
the integral is equal to the Laplace transform of $\psi_m(\tau_m,r)$. Thus
we obtain directly~\eqref{psi:compound}. 
Using the Laplace transform of the exponential distribution
$\psi_m(\tau_m,r) = \tau_k(r)^{-1} \exp[-\tau_m/\tau_k(r)]$,
\begin{align}
\hat \psi_m(\lambda,r) = \frac{1}{1 +  \lambda \tau_k(r)}
\end{align}
gives directly~\eqref{psi:c:exp}.
\section{Asymptotic Behavior\label{App:B}}
We consider now the asymptotic long time behavior of the breakthrough
curves for the CTRW models presented in
Sections~\ref{Sec:CTRW},~\ref{Sec:MRMT} and~\ref{Sec:CTRWMRMT}.
The distribution $f(t,r)$ of arrival times at a radius $r$ is given by 
\begin{align}
f(t,r) = \sum\limits_{n=0}^\infty f_0(n,r) p(t,n), 
\end{align}
where $f_0(n,r)$ is the distribution of the number of steps needed to
arrive at the radius $r$ in the random walk~\eqref{cCTRW:a}, and
$p(t,n)$ the distribution of times needed to make $n$ steps in the
random walk~\eqref{cCTRW:b}. For large
$n$ it can be approximated by an inverse Gaussian distribution that is
sharply peaked about $n = r/\ell$. Thus, we approximate the
breakthrough curve by 
\begin{align}
f(t,r) \approx p(t,r/\ell). 
\end{align}
In the following, we derive the asymptotic behavior of $p(t,n)$ for
large times $t$. Notice that $p(t,n)$ can be defined by 
\begin{align}
p(t,n) = \left \langle \delta(t - t_n) \right \rangle,
\end{align}
where $t_n$ is given by~\eqref{cCTRW:b} and can be written as 
\begin{align}
t_n = \sum\limits_{i = 0}^{n-1} \tau_i(r_i). 
\end{align}
%
Thus, the Laplace transform of $p(t,n)$ can be written as
\begin{align}
\label{B5}
\hat p(\lambda,n) = \left \langle \exp\left[- \lambda \sum\limits_{i =
  0}^{n-1} \tau_i(r_i)
  \right]  \right \rangle = \prod\limits_{i = 0}^{n-1} \hat \psi(\lambda,r_i),
\end{align}
where $\hat \psi(\lambda,r_i)$ is given by 
\begin{align}
\label{App:ps:l}
\hat \psi(\lambda,r_i) = \left \langle \exp\left[- \lambda\tau_i(r_i)
  \right]  \right \rangle.
\end{align}
Equation~\eqref{B5} can also be written as
\begin{align}
\label{B5:2}
\hat p(\lambda,n) = \exp\left\{\sum\limits_{i = 0}^{n-1} \ln[\hat
  \psi(\lambda,r_i)] \right\},
\end{align}
We note that $\hat \psi(\lambda,r_i) = 1 + \Delta \hat
\psi(\lambda,r_i)$, where $\Delta \hat \psi(\lambda,r_i)$ decreases
with decreasing $\lambda$, see below. Thus, for small $\lambda$, we
can approximate 
\begin{align}
\label{B5:3}
\hat p(\lambda,n) = \exp\left[\sum\limits_{i = 0}^{n-1} \Delta \psi(\lambda,r_i) \right],
\end{align}
The asymptotic behavior will be determined starting from
expression~\eqref{App:ps:l} for the density of a single radial transition time. 
\subsection{Heterogeneous Advection Model\label{App:B:1}}
For the CTRW model defined through~\eqref{CTRW}, the single transition
time $\tau_i(r_i)$ is defined by 
\begin{align}
\label{App:tau}
\tau_i(r_i) = \tau_k(r_i) \eta_i.
\end{align}
%
The random variable $\eta_i$ is distributed according to a power-law such that 
\begin{align}
\label{App:psieta}
\psi_\eta(\eta) \sim \eta^{-1 - \beta},
\end{align}
for large $\eta$ and $0< \beta < 2$. Thus,~\eqref{App:ps:l} can be written as 
\begin{align}
\label{App:ps:l:2}
\hat \psi(\lambda,r_i) = \left \langle \exp\left[- \lambda \tau_k(r_i)
    \eta_i \right]  \right \rangle = \hat \psi_\eta[\lambda \tau_k(r_i)].
\end{align}

Notice that the Laplace transform of the power-law
density~\eqref{App:psieta} can be expanded for small $\lambda$
as~\cite[e.g.,][]{dentz:2004} 
\begin{align}
\hat \psi_\eta(\lambda) &= 1 - \alpha_{11} \lambda^\beta, & 0 &< \beta <
1\\
\hat \psi_\eta(\lambda) &= 1 - \alpha_{12} \lambda + \alpha_{22}
\lambda^\beta, & 1 &< \beta < 2, 
\end{align}
where the parameters $\alpha_{11}$, $\alpha_{12}$ and $\alpha_{22}$ depend on the details
of the underlying distribution $\psi_\eta(\eta)$. For the Pareto
distribution~\eqref{eta:pareto}, they are given by 
\begin{align}
\alpha_{11} &= \Gamma(1 - \beta), &&
\\
\alpha_{12} &= \frac{1}{\beta -1}, && \alpha_{22} = \frac{\Gamma(2 -
  \beta)}{\beta -1}.  
\end{align}
Thus, $\hat \psi(\lambda,r_i)$ can be approximated for $\lambda \tau_k(r_i)
\ll 1$ as 
\begin{subequations}
\label{App:psim:l}
\begin{align}
\hat \psi(\lambda,r_i) &= 1 - \alpha_{11} [\lambda \tau_k(r_i)]^\beta, & 0 &< \beta <
1\\
\hat \psi(\lambda,r_i) &= 1 - \alpha_{12} \lambda \tau_k(r_i) + \alpha_{22}
[\lambda \tau_k(r_i)]^\beta, & 1 &< \beta < 2. 
\end{align}
\end{subequations}
Inserting the latter into~\eqref{B5:3} gives
\begin{align}
\label{App:p:adv1}
\hat p(\lambda,n) &= \exp\left[- \alpha_{11} (\lambda \theta_n)^\beta \right], & 0 &< \beta <
1\\
\label{App:p:adv2}
\hat p(\lambda,n) &= \exp\left[ -\lambda \langle \tau_{a,n} \rangle  + \alpha_{22}
(\lambda \theta_n)^\beta \right] , & 1 &< \beta < 2, 
\end{align}
where we defined 
\begin{align}
\theta_n = \left[\sum\limits_{i=1}^n \tau_k(r_i)^\beta
\right]^{1/\beta}, && \langle \tau_{a,n} \rangle = \alpha_{12} \sum\limits_{i=1}^n
\tau_k(r_i). 
\end{align}
The latter is the mean arrival time, which is only defined for $1 < \beta <
2$. It is impacted on by the distribution of dimensionless transition
times through $\alpha_{12}$. For $\alpha_{12} = 1$, it is equal to the mean
arrival time of the homogeneous model. From~\eqref{App:p:adv1}
and~\eqref{App:p:adv2}, we can deduce the scaling forms 
\begin{align}
\label{App:p:adv1:rt}
p(t,n) &= \frac{1}{\theta_n} f_{01}(t/\theta_n), & 0 &< \beta <
1\\
\label{App:p:adv2:rt}
p(t,n) &= \frac{1}{\theta_n} f_{02}\left(\frac{t - \langle
    \tau_{a,n}\rangle}{\theta_n} \right) , & 1 &< \beta < 2, 
\end{align}
where the Laplace transforms of $f_{01}(t)$ and $f_{02}(t)$ are
defined by 
\begin{align}
\label{App:p:adv12}
\hat f_{01}(\lambda) &= \exp\left(- \alpha_{11} \lambda^\beta \right), & 0 &< \beta <
1\\
\label{App:p:adv22}
\hat f_{02}(\lambda) &= \exp\left(\alpha_{22}\lambda^\beta \right) , & 1 &< \beta < 2, 
\end{align}

We obtain the long-time behavior of the breakthrough curves by 
expanding the latter expressions up to leading order in
$\lambda \ll 1$, which gives
\begin{align}
\hat f_{01}(\lambda) &= 1 - \alpha_{11} \lambda^\beta, & 0 &< \beta <
1\\
\hat f_{02}(\lambda) &= 1  + \alpha_{22} \lambda^\beta , & 1 &< \beta < 2. 
\end{align}
Thus, $f_{01}(t)$ and $f_{02}(t)$ behave at long times as $\sim
t^{-1-\beta}$ and thus $p(t,n)$ behaves for $t \gg \theta_n$ as 
%
\begin{align}
\label{App:pn:CTRW}
p(t,n) \sim \frac{1}{\theta_n}  \left(\frac{t}{\theta_n}\right)^{-1-\beta},
\end{align}
Expression~\eqref{fhetadv} is obtained by setting $n = r/\ell$, $r_i =
i \ell$, and $\theta(r) = \theta_{r/\ell}$. 
\subsection{MRMT Model\label{App:B:2}}
We employ for the trapping time distribution $p_{im}(\tau)$ the
power-law
\begin{align}
\label{App:pim}
p_{im}(\tau) \sim \frac{1}{\tau_c} \left(\frac{t}{\tau_c} \right)^{-1 - \delta}, 
\end{align}
for $t \gg \tau_c$, $\tau_c$ a characteristic immobile time and $0 <
\delta < 1$. Thus, its Laplace transform for $\lambda \tau_c \ll 1$
can be written as
\begin{align}
\label{App:pim:l}
\hat p_{im}(\lambda) = 1 - \alpha_{im} (\lambda \tau_c)^\delta. 
\end{align}
Inserting the latter into~\eqref{psi:c:exp} for the
transition time distribution and expanding the resulting expression up
to the leading order, we obtain
\begin{align}
\hat \psi_\tau(\lambda,r_i) =  1 - \gamma \tau_k(r_i) \alpha_{im}
(\lambda \tau_c)^\delta. 
\end{align}
Using this expression in~\eqref{B5} and again expanding to leading
order yields for $\hat p(\lambda,n)$ 
\begin{align}
\hat p(\lambda,n) = 1 - \alpha_{im}
(\lambda \tau_c)^\delta \gamma  \sum\limits_{i = 1}^n \tau_k(r_i), 
\end{align}
where $\alpha_{im}$ is given by the details of $p_{im}(\tau)$.
Thus, the asymptotic behavior of $p(t,n)$ is given by 
\begin{align}
\label{App:pn:MRMT}
p(t,n) \sim \frac{1}{\theta_{c,n}} \left(\frac{t}{\theta_{c,n}}\right)^{-1 - \delta}, &&
\theta_{c,n} = \tau_c \left[\gamma  \sum\limits_{i = 1}^n \tau_k(r_i)
\right]^{1/\delta}. 
\end{align}
%
Expression~\eqref{fMRMT} is obtained by setting $r_i = i \ell$, $n =
r / \ell$ and $\theta_c(r) = \theta_{c,r/\ell}$. 
\subsection{Heterogeneous Advection with MRMT\label{App:B:3}}
We consider now the case that the mobile transition time is given
by~\eqref{App:tau} with $\eta$ distributed according to the
power-law~\eqref{App:psieta}. For the immobile times with use the
distribution~\eqref{App:pim}. Thus, the Laplace transforms of the
distributions of the mobile transition time and the immobile times are
given by~\eqref{App:psim:l} and~\eqref{App:pim:l},
respectively. Using~\eqref{App:pim:l} in~\eqref{psi:compound} gives
for the transition time distribution
\begin{align}
\label{App:psi:compound}
\hat \psi_\tau(\lambda,r) = \hat \psi_m\left(\lambda + \gamma
  \alpha_{im} (\lambda \tau_c)^{\delta},r \right).
\end{align}
Using now expansion~\eqref{App:psim:l} for $\hat \psi_\tau(\lambda,r)$
gives 
\begin{subequations}
\label{App:psim:l:2}
\begin{align}
\hat \psi_\tau(\lambda,r_i) &= 1 - \alpha_1 \left\{\left[\lambda + \gamma
  \alpha_{im} (\lambda \tau_c)^{\delta}\right] \tau_k(r_i)\right\}^\beta, & 0 &< \beta <
1\\
\hat \psi_\tau(\lambda,r_i) &= 1 - \alpha_1 \left[\lambda + \gamma
  \alpha_{im} (\lambda \tau_c)^{\delta}\right] \tau_k(r_i) &&
\nonumber
\\
& + \alpha_2
\left\{\left[\lambda + \gamma
  \alpha_{im} (\lambda \tau_c)^{\delta}\right]
\tau_k(r_i)\right\}^\beta, & 1 &< \beta < 2. 
\end{align}
\end{subequations}
From these expressions, we can identify two characteristic times
scales that separate time regimes with different behaviors. The first
one is set by the condition 
\begin{align}
\lambda \gg \left(\alpha_{im} \gamma \tau_c\right)^{\frac{1}{1-\delta}} \tau_c^{-1},
\end{align}
which implies that 
\begin{align}
t \ll \tau_c \left(\alpha_{im} \gamma \tau_c\right)^{\frac{1}{\delta -
  1}}.
\end{align}
The second time scale is set by the condition that $\lambda \tau_k(r_i) \ll
1$, which implies 
\begin{align}
t \gg \tau_k(r_i). 
\end{align}
These scales set the time regimes $\tau_k(r_i) \ll t \ll \tau_c \left(\alpha_{im} \gamma
  \tau_c\right)^{\frac{1}{\delta - 1}}$, and $t \gg \tau_c \left(\alpha_{im} \gamma
  \tau_c\right)^{\frac{1}{\delta - 1}}$. 

In the regime $\left(\alpha_{im} \gamma
  \tau_c\right)^{\frac{1}{1-\delta}} \tau_c^{-1} \ll \lambda \ll
\tau_k(r)^{-1}$, the transition time density $\hat \psi(\lambda,r_i)$
is given in leading order by
\begin{subequations}
\label{App:psim:l:3}
\begin{align}
\hat \psi_\tau(\lambda,r_i) &= 1 - \alpha_1 \left[\lambda
  \tau_k(r_i)\right]^\beta, & 0 &< \beta < 1\\
\hat \psi_\tau(\lambda,r_i) &= 1 - \alpha_1 \lambda \tau_k(r_i) 
+ \alpha_2 \left[\lambda \tau_k(r_i)\right]^\beta, & 1 &< \beta < 2, 
\end{align}
\end{subequations}
which is identical to~\eqref{App:psim:l}. Thus, 
the breakthrough curves for $\tau_k(r) \ll t \ll \tau_c \left(\alpha_{im} \gamma
  \tau_c\right)^{\frac{1}{\delta - 1}}$ behave as in~\eqref{App:pn:CTRW}. 

In the long time regime $t \gg \tau_c \left(\alpha_{im} \gamma
  \tau_c\right)^{\frac{1}{\delta - 1}}$, we need to distinguish
between the cases $0 < \beta < 1$ and $1 < \beta < 2$. In the former
case, the leading order of $\hat \psi_\tau(\lambda,r_i)$ for $\lambda
\ll \left(\alpha_{im} \gamma \tau_c\right)^{\frac{1}{1-\delta}}
\tau_c^{-1}$ is given by 
\begin{align}
\label{App:psim:l:3:01}
\hat \psi_\tau(\lambda,r_i) &= 1 - \alpha_{1} \alpha_{im}^{\beta} \left[\gamma \tau_k(r_i) \right]^{\beta}
\left( \lambda \tau_c\right)^{\beta\delta}. 
\end{align}
Inserting this expression into~\eqref{B5} and expanding up to leading
order gives for $\hat p(\lambda,n)$
\begin{align}
\hat p(\lambda,n) = 1 - \alpha_{1} \alpha_{im}^{\beta} \left( \lambda \tau_c\right)^{\beta\delta} 
\sum\limits_{i=1}^n \left[\gamma \tau_k(r_i)
\right]^{\beta}. 
\end{align}
Thus, $p(t,n)$ behaves asymptotically as 
\begin{align}
p(t,n) \sim t^{-1 - \beta \delta}. 
\end{align}

In the case $1 < \beta < 2$, the leading order of $\hat
\psi(\lambda,r_i)$ for $\lambda
\ll \left(\alpha_{im} \gamma \tau_c\right)^{\frac{1}{1-\delta}}
\tau_c^{-1}$ is given by
\begin{align}
\label{App:psim:l:3:12}
\hat \psi(\lambda,r_i) &= 1 - \alpha_1 \alpha_{im} (\lambda
\tau_c)^{\delta} \gamma \tau_k(r_i). 
\end{align}
Inserting this expression into~\eqref{B5} and expanding up to leading
order gives for $\hat p(\lambda,n)$ similar as in the previous section
\begin{align}
\hat p(\lambda,n) = 1 - \alpha_1\alpha_{im} 
(\lambda \tau_c)^\delta \gamma  \sum\limits_{i = 1}^n \tau_k(r_i). 
\end{align}
Thus, the asymptotic behavior of $p(t,n)$ is given
by~\eqref{App:pn:MRMT}.






\bibliographystyle{elsarticle-num}

\begin{thebibliography}{10}
\expandafter\ifx\csname url\endcsname\relax
  \def\url#1{\texttt{#1}}\fi
\expandafter\ifx\csname urlprefix\endcsname\relax\def\urlprefix{URL }\fi
\expandafter\ifx\csname href\endcsname\relax
  \def\href#1#2{#2} \def\path#1{#1}\fi

\bibitem{BG1990}
J.~P. Bouchaud, A.~Georges, Anomalous diffusion in disordered media:
  {S}tatistical mechanisms, models and physical applications, Phys. Rep.
  195~(4,5) (1990) 127--293.
\newblock \href {http://dx.doi.org/10.1016/0370-1573(90)90099-N}
  {\path{doi:10.1016/0370-1573(90)90099-N}}.

\bibitem{berkowitzcortis06}
B.~Berkowitz, A.~Cortis, M.~Dentz, H.~Scher, Modeling non-{F}ickian transport
  in geological formations as a continuous time random walk, Rev. Geophys.
  44~(2) (2006) RG2003.
\newblock \href {http://dx.doi.org/10.1029/2005RG000178}
  {\path{doi:10.1029/2005RG000178}}.

\bibitem{Neuman1987}
S.~P. Neuman, C.~L. Winter, C.~M. Newman, Stochastic-theory of field scale
  {F}ickian dispersion in anisotropic porous media, Water Resour. Res. 23
  (1987) 453Ð466.
\newblock \href {http://dx.doi.org/10.1029/WR023i003p00453}
  {\path{doi:10.1029/WR023i003p00453}}.

\bibitem{CHG1994}
J.~Cushman, X.~Hu, T.~Ginn,
  \href{http://dx.doi.org/10.1007/BF02186747}{Nonequilibrium statistical
  mechanics of preasymptotic dispersion}, J. Stat. Phys. 75~(5-6) (1994)
  859--878.
\newblock \href {http://dx.doi.org/10.1007/BF02186747}
  {\path{doi:10.1007/BF02186747}}.
\newline\urlprefix\url{http://dx.doi.org/10.1007/BF02186747}

\bibitem{Benson2000}
D.~A. Benson, S.~W. Wheatcraft, M.~M. Meerschaert,
  \href{http://dx.doi.org/10.1029/2000WR900031}{Application of a fractional
  advection-dispersion equation}, Water Resour. Res. 36~(6) (2000) 1403--1412.
\newblock \href {http://dx.doi.org/10.1029/2000WR900031}
  {\path{doi:10.1029/2000WR900031}}.
\newline\urlprefix\url{http://dx.doi.org/10.1029/2000WR900031}

\bibitem{Benson2004}
D.~A. Benson, C.~Tadjeran, M.~M. Meerschaert, I.~Farnham, G.~Pohll,
  \href{http://dx.doi.org/10.1029/2004WR003314}{Radial fractional-order
  dispersion through fractured rock}, Water Resour. Res. 40~(12) (2004) W12416,
  w12416.
\newblock \href {http://dx.doi.org/10.1029/2004WR003314}
  {\path{doi:10.1029/2004WR003314}}.
\newline\urlprefix\url{http://dx.doi.org/10.1029/2004WR003314}

\bibitem{haggertyetal:1995WRR}
R.~Haggerty, S.~M. Gorelick,
  \href{http://dx.doi.org/10.1029/95WR10583}{Multiple-rate mass transfer for
  modeling diffusion and surface reactions in media with pore-scale
  heterogeneity}, Water Resour. Res. 31~(10) (1995) 2383--2400.
\newblock \href {http://dx.doi.org/10.1029/95WR10583}
  {\path{doi:10.1029/95WR10583}}.
\newline\urlprefix\url{http://dx.doi.org/10.1029/95WR10583}

\bibitem{Carrera1998}
J.~Carrera, X.~S\'anchez-Vila, I.~Benet, A.~Medina, G.~Galarza, J.~Guimer\`a,
  \href{http://dx.doi.org/10.1007/s100400050143}{On matrix diffusion:
  formulations, solution methods and qualitative effects}, Hydrogeol. J. 6~(1)
  (1998) 178--190.
\newblock \href {http://dx.doi.org/10.1007/s100400050143}
  {\path{doi:10.1007/s100400050143}}.
\newline\urlprefix\url{http://dx.doi.org/10.1007/s100400050143}

\bibitem{berkowitz1997}
B.~Berkowitz, H.~Scher,
  \href{http://link.aps.org/doi/10.1103/PhysRevLett.79.4038}{Anomalous
  transport in random fracture networks}, Phys. Rev. Lett. 79 (1997)
  4038--4041.
\newblock \href {http://dx.doi.org/10.1103/PhysRevLett.79.4038}
  {\path{doi:10.1103/PhysRevLett.79.4038}}.
\newline\urlprefix\url{http://link.aps.org/doi/10.1103/PhysRevLett.79.4038}

\bibitem{dentz:2004}
M.~Dentz, A.~Cortis, H.~Scher, B.~Berkowitz,
  \href{http://www.sciencedirect.com/science/article/pii/S0309170803001726}{Ti%
me behavior of solute transport in heterogeneous media: transition from
  anomalous to normal transport}, Adv. Water Resour. 27~(2) (2004) 155 -- 173.
\newblock \href {http://dx.doi.org/10.1016/j.advwatres.2003.11.002}
  {\path{doi:10.1016/j.advwatres.2003.11.002}}.
\newline\urlprefix\url{http://www.sciencedirect.com/science/article/pii/S03091%
70803001726}

\bibitem{Neuman_Tartakovsky2008}
S.~P. Neuman, D.~M. Tartakovsky, Perspective on theories of anomalous transport
  in heterogeneous media, Adv. Water Resour. 32 (2008) 670--680.
\newblock \href {http://dx.doi.org/10.1016/j.advwatres.2008.08.005}
  {\path{doi:10.1016/j.advwatres.2008.08.005}}.

\bibitem{dentz:2011}
M.~Dentz, T.~LeBorgne, A.~Englert, B.~Bijeljic, Mixing, spreading and reaction
  in heterogeneous media: A brief review, J. Cont. Hydrol. 120-121 (2011)
  1--17.
\newblock \href {http://dx.doi.org/10.1016/j.jconhyd.2010.05.002}
  {\path{doi:10.1016/j.jconhyd.2010.05.002}}.

\bibitem{delay_review2005}
F.~Delay, P.~Ackerer, C.~Danquigny, Simulating solute transport in porous or
  fractured formations using random walk particle tracking: A review, Vadose
  Zone Journal 4 (2005) 360--379.
\newblock \href {http://dx.doi.org/10.2136/vzj2004.0125}
  {\path{doi:10.2136/vzj2004.0125}}.

\bibitem{SalamonRW}
P.~Salamon, D.~Fern\`andez-Garcia, G.-H.~J. J., A review and numerical
  assessment of the random walk particle tracking methods, J. Cont. Hydrol. 87
  (2006) 277--305.
\newblock \href {http://dx.doi.org/10.1016/j.jconhyd.2006.05.005}
  {\path{doi:10.1016/j.jconhyd.2006.05.005}}.

\bibitem{leborgnedentz08-prl}
T.~{Le Borgne}, M.~Dentz, J.~Carrera, Lagrangian statistical model for
  transport in highly heterogeneous velocity fields, Phys. Rev. Lett. 101
  (2008) 090601.
\newblock \href {http://dx.doi.org/10.1103/PhysRevLett.101.090601}
  {\path{doi:10.1103/PhysRevLett.101.090601}}.

\bibitem{DCa:2009}
M.~Dentz, A.~Castro, Effective transport dynamics in porous media with
  heterogeneous retardation properties, Geophys. Res. Lett. 36 (2009) L03403.
\newblock \href {http://dx.doi.org/10.1029/2008GL036846}
  {\path{doi:10.1029/2008GL036846}}.

\bibitem{DentzBolster2010}
M.~Dentz, D.~Bolster, Distribution- versus correlation-induced anomalous
  transport in quenched random velocity fields, Phys. Rev. Lett. 105 (2010)
  244301.
\newblock \href {http://dx.doi.org/10.1103/PhysRevLett.105.244301}
  {\path{doi:10.1103/PhysRevLett.105.244301}}.

\bibitem{kangetal11-prl}
P.~K. Kang, M.~Dentz, T.~{Le Borgne}, R.~Juanes, Spatial {M}arkov model of
  anomalous transport through random lattice networks, Phys. Rev. Lett. 107
  (2011) 180602.
\newblock \href {http://dx.doi.org/10.1103/PhysRevLett.107.180602}
  {\path{doi:10.1103/PhysRevLett.107.180602}}.

\bibitem{Berkowitz2002}
B.~Berkowitz, J.~Klafter, R.~Metzler, H.~Scher, Physical pictures of transport
  in heterogeneous media: Advection-dispersion, random-walk, and fractional
  derivative formulations, Water Resour. Res. 38~(10) (2002) 1191.
\newblock \href {http://dx.doi.org/10.1029/2001WR001030}
  {\path{doi:10.1029/2001WR001030}}.

\bibitem{schumer:2003}
R.~Schumer, D.~A. Benson, M.~M. Meerschaert, B.~Baeumer, Fractal
  mobile/immobile solute transport, Water Resour. Res. 39 (2003) 10.
\newblock \href {http://dx.doi.org/10.1029/2003WR002141}
  {\path{doi:10.1029/2003WR002141}}.

\bibitem{dentz:2003}
M.~Dentz, B.~Berkowitz, Transport behavior of a passive solute in continuous
  time random walks and multirate mass transfer, Water Resour. Res. 39 (2003)
  1111.
\newblock \href {http://dx.doi.org/10.1029/2001WR001163}
  {\path{doi:10.1029/2001WR001163}}.

\bibitem{Margolin:et:al:2003}
G.~Margolin, M.~Dentz, B.~Berkowitz, Continuous time random walk and multirate
  mass transfer modeling of sorption, Chem. Phys. 295 (2003) 71--80.
\newblock \href {http://dx.doi.org/10.1016/j.chemphys.2003.08.007}
  {\path{doi:10.1016/j.chemphys.2003.08.007}}.

\bibitem{BensonMRMT2009}
D.~A. Benson, M.~M. Meerschaert, A simple and efficient random walk solution of
  multi-rate mobile/immobile mass transport equations, Adv. Water Resour. 32
  (2009) 532Ð539.
\newblock \href {http://dx.doi.org/10.1016/j.advwatres.2009.01.002}
  {\path{doi:10.1016/j.advwatres.2009.01.002}}.

\bibitem{DentzGouzeAWR2012}
M.~Dentz, P.~Gouze, A.~Russian, J.~Dweik, F.~Delay, Diffusion and trapping in
  heterogeneous media: {A}n inhomogeneous continuous time random walk approach,
  Adv. Water Resour. 49 (2012) 13--22.
\newblock \href {http://dx.doi.org/10.1016/j.advwatres.2012.07.015}
  {\path{doi:10.1016/j.advwatres.2012.07.015}}.

\bibitem{LeBorgneGouze2008}
T.~Le~Borgne, P.~Gouze, Non-{F}ickian dispersion in porous media: 2. model
  validation from measurements at different scales., Water Resour. Res. 44
  (2007) W06427.
\newblock \href {http://dx.doi.org/10.1029/2007WR006279}
  {\path{doi:10.1029/2007WR006279}}.

\bibitem{Benson:2008}
Y.~Zhang, D.~A. Benson, Lagrangian simulation of multidimensional anomalous
  transport at the made site, Geophys. Res. Lett. 35 (2008) L07403.
\newblock \href {http://dx.doi.org/10.1029/2008GL033222}
  {\path{doi:10.1029/2008GL033222}}.

\bibitem{Haggerty:2001WRR}
R.~Haggerty, S.~W. Fleming, L.~C. Meigs, S.~A. McKenna, Tracer tests in a
  fractured dolomite 2. analysis of mass transfer in single-well
  injection-withdrawal tests, Water Resour. Res. 37 (2001) 1129--1142.
\newblock \href {http://dx.doi.org/10.1029/2000WR900334}
  {\path{doi:10.1029/2000WR900334}}.

\bibitem{Becker:2003wd}
M.~W. Becker, A.~M. Shapiro, Interpreting tracer breakthrough tailing from
  different forced- gradient tracer experiment configurations in fractured
  bedrock, Water Resour. Res. 39 (2003) 1024.
\newblock \href {http://dx.doi.org/10.1029/2001WR001190}
  {\path{doi:10.1029/2001WR001190}}.

\bibitem{Kang2015}
P.~K. Kang, T.~Le~Borgne, M.~Dentz, O.~Bour, R.~Juanes, Impact of velocity
  correlation and distribution on transport in fractured media: field evidence
  and theoretical model, Water Resour. Res. 51 (2015) 940--959.
\newblock \href {http://dx.doi.org/10.1002/ 2014WR015799} {\path{doi:10.1002/
  2014WR015799}}.

\bibitem{Risken:1996}
H.~Risken, The Fokker-Planck Equation, Springer Heidelberg New York, 1996.

\bibitem{Bear:1972}
J.~Bear, Dynamics of fluids in porous media, American Elsevier, New York, 1972.

\bibitem{AS1972}
M.~Abramowitz, I.~A. Stegun, Handbook of Mathematical Functions, Dover
  Publications, New York, 1972.

\bibitem{silva:2009}
O.~Silva, J.~Carrera, D.~M., S.~Kumar, A.~Alcolea, M.~Willmann, A general
  real-time formulation for multi-rate mass transfer problems, Hydrol. Earth
  Syst. Sci. 13 (2009) 1399--1411.
\newblock \href {http://dx.doi.org/10.5194/hess-13-1399-2009}
  {\path{doi:10.5194/hess-13-1399-2009}}.

\bibitem{Edery2014}
Y.~Edery, A.~Guadagnini, H.~Scher, B.~Berkowitz, Origins of anomalous transport
  in heterogeneous media: Structural and dynamic controls, Water Resour. Res.
  50.
\newblock \href {http://dx.doi.org/10.1002/2013WR015111}
  {\path{doi:10.1002/2013WR015111}}.

\bibitem{HARV95}
C.~F. Harvey, S.~M. Gorelick, Temporal moment-generating equations: Modeling
  transport and mass transfer in heterogeneous aquifers, Water Resour. Res.
  31~(8) (1995) 1895--1911.
\newblock \href {http://dx.doi.org/10.1029/95WR01231}
  {\path{doi:10.1029/95WR01231}}.

\bibitem{Tecklenburg2013}
J.~Tecklenburg, I.~Neuweiler, M.~Dentz, J.~Carrera, S.~Geiger, C.~Abramowski,
  O.~Silva, A non-local two-phase flow model for immiscible displacement in
  highly heterogeneous porous media and its parametrization, Adv. Water Resour.
  62 (2013) 475Ð487.
\newblock \href {http://dx.doi.org/10.1016/j.advwatres.2013.05.012}
  {\path{doi:10.1016/j.advwatres.2013.05.012}}.

\bibitem{DGC2011}
M.~Dentz, P.~Gouze, J.~Carrera, Effective non-local reaction kinetics for
  transport in physically and chemically heterogeneous media, J. Cont. Hydrol.
  120-121 (2011) 222Ð236.
\newblock \href {http://dx.doi.org/10.1016/j.jconhyd.2010.06.002}
  {\path{doi:10.1016/j.jconhyd.2010.06.002}}.

\bibitem{Noetinger:2000}
B.~Noetinger, T.~Estebenet, Up-scaling of double porosity fractured media using
  continuous-time random walks methods, Transp. Porous Media 39 (2000)
  315--337.
\newblock \href {http://dx.doi.org/10.1023/A:1006639025910}
  {\path{doi:10.1023/A:1006639025910}}.

\bibitem{GMDC2008}
P.~Gouze, Z.~Melean, T.~Le~Borgne, M.~Dentz, J.~Carrera, Non-fickian dispersion
  in porous media explained by heterogeneous microscale matrix diffusion, Water
  Resour. Res. 44 (2008) W11416.
\newblock \href {http://dx.doi.org/10.1029/2007WR006690}
  {\path{doi:10.1029/2007WR006690}}.

\bibitem{Willmann2008}
M.~Willmann, J.~Carrera, X.~Sanchez-Vila, Transport upscaling in heterogeneous
  aquifers: {W}hat physical parameters control memory functions?, Water Resour.
  Res. 44 (2008) W12437.
\newblock \href {http://dx.doi.org/10.1029/2007WR006531}
  {\path{doi:10.1029/2007WR006531}}.

\bibitem{Zhang2014}
Y.~Zhang, C.~T. Green, B.~Baeumer, Linking aquifer spatial properties and
  non-{Fickian} transport in mobile--immobile like alluvial settings, J.
  Hydrol. 512 (2014) 315--331.
\newblock \href {http://dx.doi.org/10.1016/j.jhydrol.2014.02.064}
  {\path{doi:10.1016/j.jhydrol.2014.02.064}}.

\end{thebibliography}

\end{document}